\documentclass[10pt,letterpaper]{article}
\usepackage[top=0.85in,left=2.75in,footskip=0.75in]{geometry}

\usepackage{amsmath,amssymb}

\usepackage{changepage}

\usepackage[utf8x]{inputenc}

\usepackage{textcomp,marvosym}

\usepackage{cite}

\usepackage{nameref,hyperref}
\usepackage{subcaption}
\usepackage[right]{lineno}

\usepackage{microtype}
\DisableLigatures[f]{encoding = *, family = * }

\usepackage[table]{xcolor}

\usepackage{array}

\usepackage{graphicx}

\usepackage{epstopdf}
\AppendGraphicsExtensions{.tiff}

\newcolumntype{+}{!{\vrule width 2pt}}

\newlength\savedwidth

\raggedright
\usepackage[aboveskip=1pt,labelfont=bf,labelsep=period,justification=raggedright,singlelinecheck=off]{caption}

\makeatletter
\renewcommand{\@biblabel}[1]{\quad#1.}
\makeatother

\usepackage{lastpage,fancyhdr,graphicx}
\usepackage{epstopdf}
\fancyheadoffset[L]{2.25in}
\fancyfootoffset[L]{2.25in}
\begin{document}
\vspace*{0.2in}

\begin{flushleft}
{\Large
\textbf\newline{Unsupervised machine learning framework for discriminating major variants of concern  during COVID-19}  
}
\newline 
\\
Rohitash Chandra\textsuperscript{1*},
Chaarvi Bansal\textsuperscript{1,3},
Mingyue Kang\textsuperscript{1},
Tom Blau\textsuperscript{4*},
Vinti Agarwal\textsuperscript{3},
Pranjal Singh\textsuperscript{5},
Laurence O. W. Wilson\textsuperscript{2},
Seshadri Vasan\textsuperscript{6}

\bigskip
\textbf{1} Transitional Artificial Intelligence Research Group, School of Mathematics and Statistics, UNSW Sydney, Sydney, Australia
\\
\textbf{2} Australian e-Health Research Centre, Commonwealth Scientific and Industrial Research Organisation, North Ryde, Australia
\\
\textbf{3} Department of Computer Science and Information Systems, Birla Institute of Technology and Science Pilani, Rajasthan, India
\\
\textbf{4} Data 61, CSIRO, Sydney, Australia
\\
\textbf{5} Department of Computer Science and Engineering, Indian Institute of Technology Guwathi, Assam,  India
\\
\textbf{6} Department of Health Sciences, University of York, York,  United Kingdom
\\
\bigskip 

* rohitash.chandra@unsw.edu.au\\
* tom.blau@data61.csiro.au

\end{flushleft}
\section*{Abstract}

 Due to the high mutation rate of the virus, the COVID-19 pandemic evolved rapidly. Certain variants of the virus, such as   Delta  and Omicron, emerged with altered viral properties leading to  severe  transmission and death rates. These variants burdened the medical systems worldwide  with  a major impact to travel, productivity, and  the world economy. Unsupervised machine learning methods have the ability to compress, characterize, and visualize unlabelled data. This paper presents a framework that utilizes unsupervised machine learning methods to discriminate and visualize the associations between major COVID-19 variants based on their genome sequences.  These methods comprise a combination of selected dimensionality reduction and clustering techniques. The framework processes the RNA sequences by performing a $k$-mer analysis on the data and further visualises and  compares the results using selected  dimensionality reduction methods that include \textit{principal component analysis} (PCA), \textit{t-distributed stochastic neighbour embedding} (t-SNE), and \textit{uniform manifold approximation projection} (UMAP). Our framework also employs agglomerative hierarchical clustering   to visualize the mutational differences among major variants of concern and country-wise mutational differences for selected  variants (Delta and Omicron) using dendrograms. We also provide country-wise mutational differences for selected variants via dendrograms. We find that the proposed framework can effectively distinguish between the major variants and has the potential to identify emerging variants in the future.

%\linenumbers
 
\section*{Introduction}

Coronaviruses (CoVs) consist of enclosed, positive-sense, single-stranded, and diversified Ribonucleic acid (RNA) viruses \cite{rna}. CoVs comprise  major variants that occur through mutations, also known as \textit{genera}, including delta, gamma, beta and alpha \cite{coronavirusgenera,coronavirus2}.  
\textcolor{black}{Currently, there are three reported highly deadly coronaviruses,   MERS-CoV, SARS-CoV-2 and SARS-CoV, due to their lethal effects on humans \cite{deadly_hcov, mers_sars}. In contrast to other CoVs, these three are more likely to cause acute lung injury, multiple organ failure and even death \cite{organ_failure}. Even though  human coronaviruses (hCoVs) \cite{humanbarrier,hcov} primarily lead  to asymptomatic or mild infections, they  cause around 15 to 30\% of common colds \cite{flu}. In 2020 the world began to witness the pandemic caused by SARS-CoV-2  which led  the detrimental repercussions to the world economy \cite{impact,sgeconomy}.   }

MERS-CoV surfaced ten years after SARS-CoV  was initially reported in April 2012 in Jordan  which accounted for the continuous outbreaks in the  Middle East region \cite{spread}. After the plague caused by MERS-CoV, SARS-CoV-2 also known as COVID-19 was first reported in Wuhan, China in December 2019. This escalated across other cities in China and threatened the health of people worldwide. On 30th January 2020, COVID-19 was declared a global concern and subsequently declared a global pandemic \cite{origin}. Notably, the SARS-CoV-2 caused mutations in humans which led to further worldwide outbreaks \cite{variation}. Although  some vaccines such as \textit{Oxford-AstraZeneca} and \textit{Pfizer-BioNTech} have been reported to limit transmission  and hospitalization rate \cite{mahase2021delta,moore2021vaccination}. SARS-CoV-2 presently does not have  fully effective vaccines.  It is imperative to efficaciously trace the virus by performing \textit{polymerase chain reaction} (PCR) \cite{tahamtan2020real} tests to sequence the strand in suspected patients. PCR tests with timely reporting can examine patterns in mutation and forecast transmission routes; however, there have been a number of challenges since the efficacy of the tests has also been questioned \cite{arevalo2020false,floriano2020accuracy}.   Hence, there is a dire need to acquire more knowledge of these two deadly hCoVs, and  combat outbreaks given emerging variants  \cite{vaccine}.

Traditionally, \textit{principal component analysis} (PCA) \cite{abdi2010principal} has been used extensively in biology to examine genome and protein sequences   to reduce the dimensions of complex datasets such as \textit{deoxyribonucleic acid} (DNA) and \textit{ribonucleic acid}  (RNA) sequences \cite{wold1993dna,eltzner2018torus}. A high dimensional dataset is projected using PCA into an \textit{eigenspace} that constitutes the direction of the largest variation illustrated by principal components. There are various drawbacks when using PCA, including the existence of abnormality that can lead to a recalculation of the PCA and result in unnecessary information disclosure \cite{pcaproblem}. Other than PCA, \textit{t-distributed stochastic neighbour embedding} (t-SNE) \cite{van2008visualizing} is widely used in the field of bioinformatics \cite{cieslak2020t}. t-SNE is capable of displaying local structure by reducing the dimensions of data; however, inaccurate representation of the global structure in the data has been reported in some studies \cite{tsneproblem}. In order to mitigate these  problems, other  approaches  such as  \textit{uniform manifold approximation projection} (UMAP)\cite{mcinnes2018umap} have been used which outperformed PCA and t-SNE for \textit{transcriptomic} datasets  \cite{comparison}. \textcolor{black}{We note that a similar study was done during the beginning of the COVID-19 pandemic \cite{bauer2020supporting}; however, only the early form of the variants, without a comprehensive evaluation of the different dimensionality reduction and clustering methods which is a major contribution of this study.}

$k$-mer analysis is typically used within the context of computational genomics and sequence analysis \cite{chor2009genomic,mapleson2017kat}. It has also been used in the analysis of COVID-19 sequences \cite{ali2021k}.
$k$-mers are sub-strings of length $k$ contained within a larger biological sequence, where the $k$-mers consist  of \textit{nucleotides} \cite{fivser201420} (i.e. A, T, G, and C).  It is important to use the right value of $k$, taking into account that larger values of $k$ increase the sequence processing time exponentially. 
 
 In this paper, we present an unsupervised machine learning framework that utilizes $k$-mer analysis for feature extraction from the selected genome (SARS-CoV-2) isolates and compares different dimensionality reduction methods that include  PCA, t-SNE, and  UMAP to visualise major variants. Furthermore, the framework employs selected clustering methods and provides a visualisation using a \textit{dendrogram} plot. First, we investigate an optimal value of $k$ for $k$-mer analysis and then evaluate the selected dimensionality reduction methods. After this, we apply \textit{agglomerative hierarchical clustering} and visualise mutational differences between variations of concern and country-wise mutational differences for selected variants via dendrograms. Our study deals with the RNA sequences of coronavirus, and hence  we use $k$-mer analysis before applying selected dimensionality reduction   and clustering methods.  We also provide an open-source code framework developed in Python  to further extend the study to emerging variants. \textcolor{black}{We investigate the effect of the prominent dimensional reduction methods  since they have certain strengths and limitations which have mostly been shown for tabular data. Hence, our major contribution is in evaluating the dimensional reduction methods with a study of visualisation produced by them for genome analysis.}

We organise the remaining sections  of this paper  as follows.  Section 2 provides an overview of the  framework  via unsupervised machine learning for distinguishing major variants. Section 3  presents the results, and Section 4 discusses the results. Lastly, Section 5 provides a conclusion of the study.

\section*{Materials and methods}

\subsection*{Data}

Nowadays, the \textit{global initiative on sharing Avian influenza} data (GISAID) \cite{gisaid} is recognized as a reliable portal for prompt sharing of COVID-19 data \cite{platform}. Currently, GISAID is the largest publicly accessible platform, consisting of sequences and associated epidemiological data of over 12.1 million SARS-CoV-2 strains (\url{https://www.gisaid.org/hcov19-variants/}). Due to the tremendous effort by scientists,  new SARS-CoV-2 variants of concern have been  included in GISAD, such as B.1.1.7 (Alpha; first detected in the United Kingdom), B.1.617.2 (Delta; first detected in India) and B.1.1.529 (Omicron; first detected in South Africa) \cite{delta, omicron2021}. GISAID provides prompt updates to formulate crucial public health policies to control COVID-19 situations globally. 

We extracted 250 randomly selected SARS-CoV-2 isolates of complete genome sequences of human origins  from GISAID  on 12th September 2022. The five variants (Alpha, Beta, Gamma, Delta, and Omicron) featured 50 genome sequences each. Table \ref{tab:var} presents the meta-information from the top 10 countries  based on the number of   genome isolates  for the selected variants across the globe. 

\begin{table}
\small
\begin{tabular}{  l l l l  l } 
 \hline
Country & Number of Occurrences & Number of Variants \\
\hline

%MERS & MERS & 2012  & ongoing  &       \\ 
%\hline
United States & Alpha(18), Beta(5), Delta\\
& (5),Gamma(8), Omicron(5)  & 5\\
India & Alpha(10), Beta(3), Gamma\\
& (3), Delta(11), Omicron(10)   & 5\\
Brazil & Alpha(5), Beta(3), Gamma\\
& (4), Delta(5), Omicron(3)   & 5\\
Italy & Beta(5), Omicron(4) \\
& Gamma(5)  & 3\\
Japan & Alpha(7), Gamma(5) \\
& Delta(1) & 3 \\
South Africa & Alpha(2), Beta(2), Delta\\
& (3), Gamma(3), Omicron(4) & 5 \\
Poland & Delta(5), Omicron(5) & 2 \\
Canada & Alpha(2), Beta(5)\\
& Gamma(2) & 3 \\
Spain & Beta(5), Gamma(3) & 2\\
England & Omicron(7) & 1\\
 \hline
 %\end{center}
\end{tabular}

    \caption{Dataset featuring top 10 countries with the number of randomly extracted genome isolates (in brackets) based on variants across the globe. Note that our dataset features 34 countries. }
    \label{tab:var}
\end{table}

In addition, we extracted 250 further genome sequences each for Delta and Omicron on 16th September 2022 from GISAID to visualize the country-wise mutational differences  with meta-information in Table \ref{tab:omi} and Table \ref{tab:delta}.

\begin{table}
\small
\begin{tabular}{  l l l l  l } 
%\begin{center}
 \hline
Country & Number of Occurrences \\
\hline
%MERS & MERS & 2012  & ongoing  &       \\ 
%\hline
France & 42\\
South Africa & 41\\
USA & 25\\
India & 25\\
Brunei & 23\\
England & 23\\
Spain & 20\\
Denmark & 18\\
Peru & 15\\
Canada & 15\\
 \hline
 %\end{center}
\end{tabular}
    \caption{Dataset featuring top 10 countries with the number of randomly extracted genome isolates of Omicron variant. Note that our dataset features 17 countries. }
    \label{tab:omi}
\end{table}

\begin{table}
\small
\begin{tabular}{  l l l l  l } 
%\begin{center}

 \hline
Country & Number of Occurrences \\
\hline

%MERS & MERS & 2012  & ongoing  &       \\ 
%\hline
India & 83\\
USA & 47\\
France & 32\\
Denmark & 30\\
Germany & 26\\
Brazil & 18\\
Indonesia & 16\\
Italy & 15\\
Mongolia & 10\\
Sudan & 6\\

 \hline
 %\end{center}
\end{tabular}

    \caption{Dataset featuring top 10 countries with the number of randomly extracted genome isolates of Delta variant. Note that our  dataset features 17 countries. }
    \label{tab:delta}
\end{table}

%xxxxxx

\subsection*{k-mer Analysis} 

Data pre-processing methods such as $k$-mer analysis have been  prominent in the analysis of genome (DNA) sequences. $k$-mers are substrings of genome sequences of length  $k$ and analysis  is done to calculate the frequency of  the substrings. A $k$-mer refers to all of a sequence's substring of length $k$; for instance, the sequence ``ATGG" would have four monomers (A, T, G, and G), three 2-mers (AT, TG, GG), two 3-mers (ATG and TGG), and one 4-mer (ATGG). Effective $k$-mer analysis  can reduce computational time for sequence processing and provide better data storage for further analysis with statistical methods  \cite{kmer_storage}.     $k$-mer analysis is extensively used in numerous bioinformatics problems, including computational genomics and sequence analysis \cite{kmer_application} and has also been applied for COVID-19. The major challenge of $k$-mer analysis is in determining the value of ``k" which needs to be determined experimentally for different problems.  A number of packages in languages such as R and Python exist for $k$-mer analysis \cite{lorenzi2020imoka,crusoe2015khmer}. Typically, $k$-mers consisting of ambiguous bases i.e. bases not identified during sequencing, such as `N' which represents any possible nucleotide, are  deleted. After $k$-mer analysis, the  distance between a pair or a group of sequences can be visualized using unsupervised machine learning methods.

 \textcolor{black}{DNA is represented with bases paired on the opposite strands (double-stranded) \cite{khanna2001dna} and typically sequenced  on either of the two strands. We need to consider every location of the genome once, no matter  which has been considered.  For instance, our analysis for  sequence ``ATCGAC" would consider its reverse complement ``GTCGAT". In canonical $k$-mer count, the $k$-mers that are reverse complements of themselves are counted twice. Typically, $k$-mer counting tools \cite{manekar2018benchmark,lorenzi2020imoka,crusoe2015khmer}  either count in canonical $k$-mers or have the option to switch between canonical and non-canonical \cite{chor2009genomic}. We note that there are three types of $k$-mer count, which include \textit{total}, \textit{unique}, and \textit{distinct} $k$-mers.   Hence,  distinct $k$-mers would be counted only once,   while unique $k$-mers are those that appear only once.    Therefore, the sequence ``ATCGATCAC" in non-canonical form, would have 7 total 3-mers, 6 unique 3-mers, and 5 distinct 3-mers, respectively. We used the \textit{kmer package} \cite{kmer} in R with default values that used total count in con-canonical form. }

%xxxxx
\subsection*{Dimensionality Reduction}

\subsubsection*{PCA}

PCA is a dimensionality reduction  method extensively used in various forms of data reduction, data analysis, and data visualization with applications in computer graphics \cite{pca}, machine learning \cite{howley2005effect}, and bioinformatics \cite{wold1993dna,eltzner2018torus}. The aim of PCA is to calculate the most relevant linear basis to represent a complex data set.   Thus, PCA is a linear combination of the basis vectors which reduces the dimensions while retaining the most crucial information. Another assumption of PCA is that the principal components are orthogonal. This assumption is essential as it serves as an intuitive simplification which means PCA can function with linear algebra decomposition approaches. In the field of medicine, PCA is used to solve various problems, including multicollinearity clinical studies \cite{pcamed}. PCA has been used to detect phenotypes in order to forecast the severity of COVID-19 and implement an individual treatment \cite{covidpca}. Similarly, PCA has been utilized to automatically classify five types of electrocardiogram (ECG) to detect aberrant cardiac electrical activity \cite{pcaheart}. However, limitations of PCA exist in sparse   datasets, datasets with uncorrelated features, and datasets with  outliers  \cite{pcaissue}. 

\subsubsection*{t-SNE}

t-SNE is a nonlinear dimensionality reduction method that is also used for the visualization of  high-dimensional data into a low-dimensional space of two or three dimensions. t-SNE is an extension of \textit{stochastic neighbour embedding}(SNE)  \cite{hinton2002stochastic} with two key modifications that include a  student t-distribution rather than a Gaussian and a symmetrical form of the SNE cost function with basic gradients.  t-SNE has been  widely used in the domain of medicine, and bioinformatics \cite{tsneapplication} e.g. in molecular dynamics simulations of macromolecules for   visualization \cite{tsnemedicine}, and  motor behaviour in Parkinson's disease   \cite{tsnemedicine_pd}. However, a major limitation of t-SNE  is the visualization  of the entire structure of the data and the lack of information, such as explained variance ratio that is given by PCA. Since the dimensionality reduction in t-SNE is based on local properties of the data, it could face challenges in  high dimensional structure. Hence, it is important to evaluate its performance for different applications. Therefore, in this study, we compare t-SNE with other dimensionality reduction methods. 

\subsubsection*{UMAP}

UMAP is a manifold learning approach for dimensionality reduction which employs a conceptual structure according to the Riemannian geometry and algebraic topology \cite{mcinnes2018umap}. UMAP has been shown to perform comparably to t-SNE in terms of visualization quality \cite{kobak2021initialization} and potentially retains better global structure with less computation time. Additionally, UMAP does not have computational restrictions on the dimension of embedding, making it practical as a dimension reduction approach for various problems. UMAP can  be expressed in the form of weighted graphs, which places UMAP in the category of k-neighbour-based graph learning models such as \textit{Isomap} \cite{tenenbaum2000global} and t-SNE. 
Together with various k-neighbour graph-based models, UMAP can be expressed in two parts. In the first part, a specific weighted k-neighbour graph is generated, and in the second part, a low-dimensional outline of this graph is calculated. UMAP has been successful in bioinformatics problems such as  dimensionality reduction and visualization of single-cell data \cite{becht2019dimensionality} and transcriptomics data \cite{yang2021dimensionality}.

\subsection*{Agglomerative clustering } 

\textit{Hierarchical agglomerative clustering} \cite{ac_common}, also known as \textit{agglomerative nesting} (AGNES) provides a better approach by addressing the problem of $k$-means clustering, where $k$ needs to be manually tuned. In an agglomerative clustering model, the clustering initiates with individual collections of every data point \cite{ac_algorithm}. AGNES has been extensively used in various medical domains \cite{ac_domain, ac_domain2}, such as  categorizing  patients with severe aortic stenosis   \cite{ac_application}, and  mapping molecular substructures \cite{ac_application2}. However, AGNES has been  ineffective in some problems since  finding the nearest pair of clusters can be challenging when data is sparse and noisy \cite{ac_issue}.

AGNES produces a  dendrogram that visualizes the hierarchical relationship amongst the clusters. A dendrogram consists of a tree-like structure for interpretive machine learning which provides a visualisation of how the data instances are allocated  to the respective clusters.   \textit{Phylogenetic} associations interpreted from genome sequences are conventionally presented as trees, which can also be represented using    dendrograms   \cite{phylogenetic1}.

\subsection*{Framework}

\begin{figure}
    \centering
    \includegraphics[scale = 0.50]{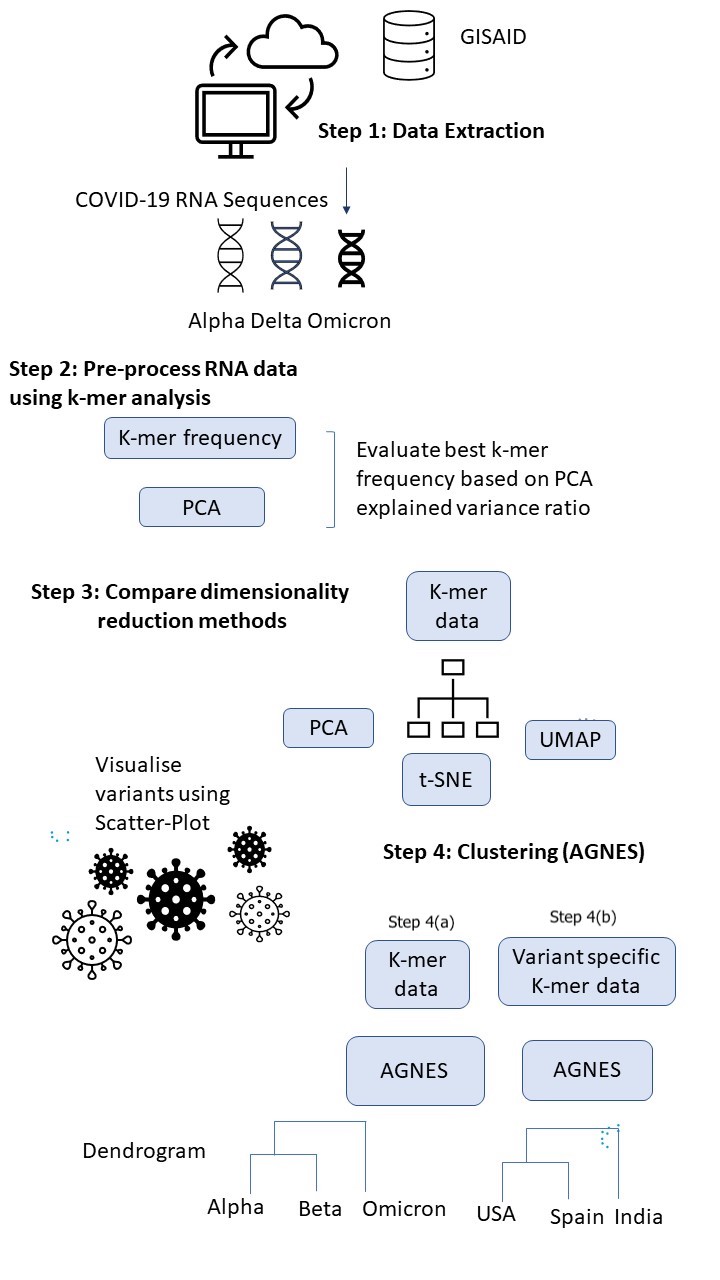}
    \vskip -0.3in
    \caption{Framework showing the major steps for analysis of COVID-19 genome sequences of major variants.}
    \label{fig:futurwork}
\end{figure}

Figure \ref{fig:futurwork}  presents the framework for discriminating and visualizing major COVID-19 variants based on genome (RNA) data of the virus. In the first step, we extract data from the  GISAID  database where we take random samples of selected variants to demonstrate the effectiveness of the framework.

In the second step, we break down the genomes into $k$-mers with selected values of $k$ and evaluate the most appropriate for effective visualization via PCA in the next step.  We remove any \textit{ambiguous base} in the genome  accordingly using the package employed by the framework \cite{kmer}. We select the best value of $k$ in $k$-mer analysis based on the \textit{explained variance ratio} of the first two \textit{principal components} of the reduced dataset. We choose the $k$ that provides the highest value of the combined explained variance ratio. The framework also reports a \textit{scree-plot} to show the explained variance ratio so that the number of principal components in PCA adequately represents the original data that can be selected. 

Subsequently, in step three, we compare the selected  dimensionality reduction approaches that include PCA, t-SNE, and UMAP. Note that our framework is general and other dimensional reduction approaches such as Isomap and linear discriminant analysis (LDA) can also be utilized. In this step, we compare the visualization produced by the first two components of the respective approaches for the selected COVID-19 variants. 

In Step 4(a), we take the data after $k$-mer analysis and apply clustering via AGNES.     We first visualize the mutational differences among the five variants (Alpha, Beta, Gamma, Delta, and Omicron), and then visualize the country-wise differences between the genome sequences of Delta and Omicron. 
 
Finally, in Step 4(b), we investigate how the variants compare with others based on their country. The major motivation  of this investigation is to track future variants as they are moving from country to country at different times.

\subsection*{Implementation}

In our proposed framework, we implement $k$-mer analysis using the  k-mer~\cite{kmer} R package and the scikit-learn Python package~\cite{scikit-learn} for implementing the dimensional reduction methods (PCA, UMAP, t-SNE). We also use the same package to implement the clustering approach and provide visualizations using standard R libraries (ggplots). Our framework is available via the GitHub repository which is included in the data section of this paper. In our experiments, we  use  the Macintosh Operating System  with an \textit{Apple M1 chip} featuring 8‑core GPU (graphics processing units) and 8‑core CPU (central processing units). Note that our framework excludes GPU and utilizes CPU computational power only.

% Results and Discussion can be combined.
\section*{Results}
\subsection*{k-mer and PCA analysis}

We first investigate the optimal  value of $k$ for  $k$-mer analysis of the selected  genomes via explained variance ratio of PCA (Step 3 of framework given in Figure \ref{fig:futurwork}). In this way, we understand the best value  obtained by different $k$-mer analyses, where $k\in\{3,5,7\}$. We  use the dataset of  250 randomly selected coronavirus sequences (Table \ref{tab:var}) for the five variants. 

\begin{figure}[htbp!]
    \centering
    \includegraphics[scale = 0.3]{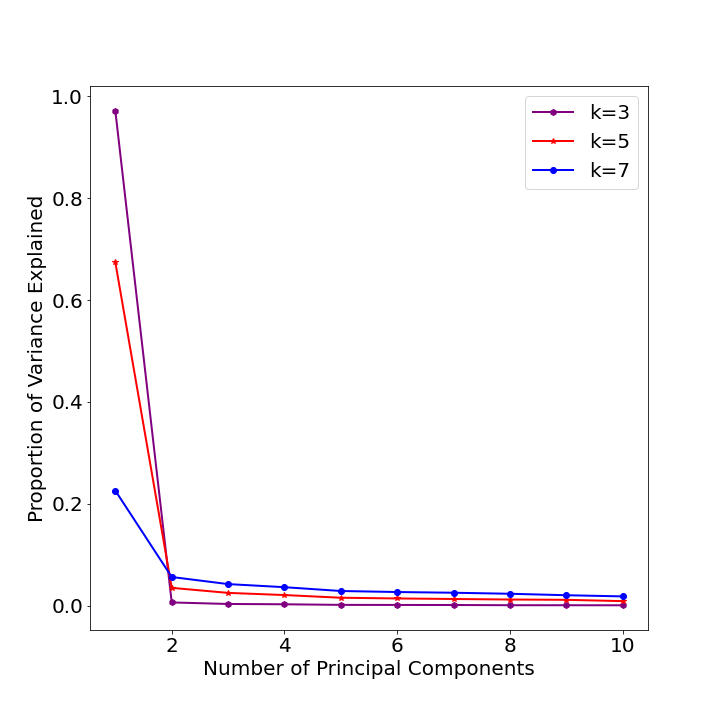}
    \caption{Scree-plot outlining the proportion of explained variance by each principal component in PCA for $k$-mer size   of 3 (purple), 5 (Red), and 7 (black).}
    \label{fig:screeplot}
\end{figure}

Figure \ref{fig:screeplot} presents the scree-plot of the proportion of variance explained by the different number of principal components (PCs) obtained via PCA for different values of $k$ in $k$-mer analysis. We observe that the total explained variance  decreases as the value of $k$ increases; hence, the best value is given by $k=3$.

Table  \ref{tab:pca} shows the top 5 principal components (PC) variance ratio for 3 selected $k$ values. Note that $k=3$ shows the highest total variance ratio; hence, this is selected for future analysis. The proportion of explained variance by the first component  is $53.3\%$ for $k=3$ with a total of around $75\%$; however, for $k=5$, the explained variance falls drastically to a total of around $44\%$. Similarly,  when $k=7$, the proportion of explained variance decreases further to  around 18\%. This means that the $k$-mer analysis with increasing values of $k$  has an inverse relationship with the  explained variance ratio.  

\begin{table}[htbp!]
    \centering
    \small
\begin{tabular}{  c c c c c c c } 
%\begin{center}
 \hline
&PC1 & PC2 & PC3 & PC4 & PC5 & Total\\
\hline
k = 3 & 0.5330 & 0.0774 & 0.0569 & 0.0519 & 0.0352 & 0.7544 \\ 
\hline
k = 5 & 0.1690 & 0.0881 & 0.0670 & 0.0617 & 0.0538 & 0.4396 \\ 
\hline
k = 7 & 0.06498 & 0.03754 & 0.0298 & 0.0258 & 0.0233 & 0.1814 \\ 
 \hline
 %\end{center}
\end{tabular}

    \caption{Explained variance ratio of top 5 principal components (PC) for selected values of $k$ in $k$-mer analysis. }
    \label{tab:pca}
\end{table}

\subsection*{Visualisation using dimensionality reduction methods}

In the previous section, we ran PCA-based dimensionality reduction to evaluate the value of $k$ in $k$-mer analysis based on explained variance ratio. Dimensionality reduction methods such as PCA can be used to visualize data via scatter plots of the first two principal components. In this way, we have a better picture of the data, giving more insight than explained variance ratio. Next, we take the same dataset, i.e., SARS-CoV-2 genome isolates from 5 distinct clusters (Table \ref{tab:var}), and run PCA and two other dimensional reduction methods (t-SNE and UMAP),  as outlined in our framework shown in Figure \ref{fig:futurwork}. We visualize the different dimensionality reduction methods by  varying  the value of $k$ and present a two-dimensional scatter plot of the first two components. Unlike PCA, t-SNE and UMAP do not provide explained variance ratio, so it is unclear what percentage of data is represented by the first two components; however, we can visually evaluate them based on the scatter plot. 
 
\begin{figure}[htbp!]
\centering
\begin{subfigure}{0.75\textwidth}
    \vskip -0.1in
    \includegraphics[width=\textwidth]{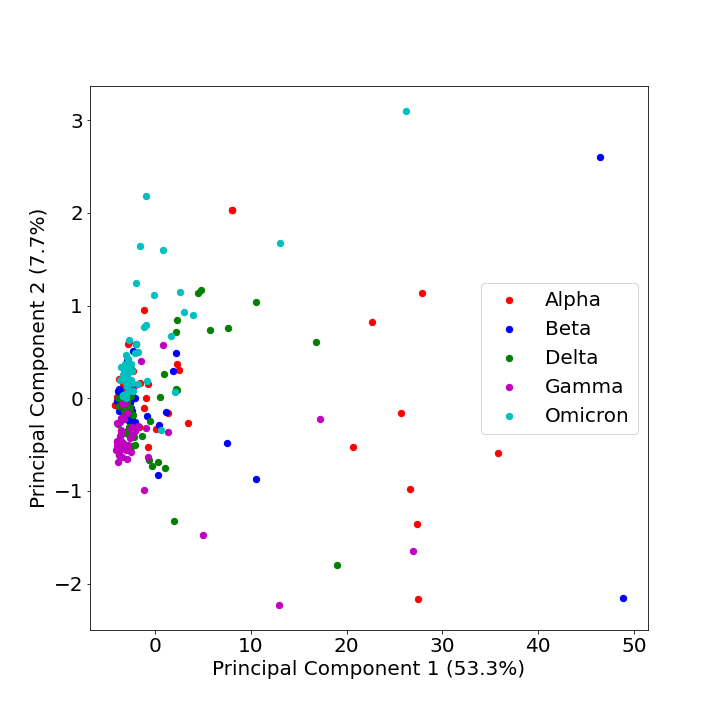}
    \vskip -0.2in
    \caption{PCA visualization }
    \label{fig:pca_covid}
\end{subfigure}
\hfill
\begin{subfigure}{0.75\textwidth}
    \includegraphics[width=\textwidth]{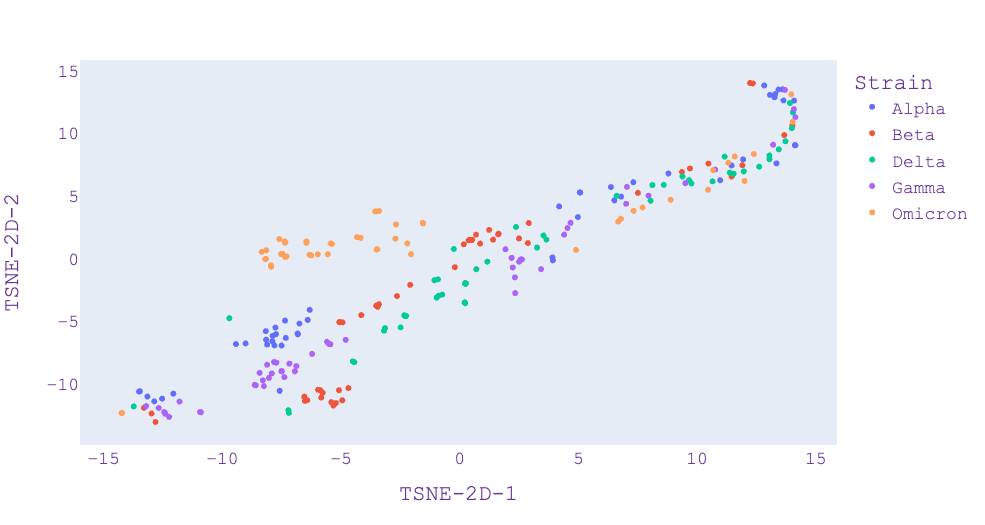}
    \vskip -0.1in
    \caption{UMAP visualization }
    \label{fig:umap}
\end{subfigure}
\hfill
\begin{subfigure}{0.75\textwidth}
    \includegraphics[width=\textwidth]{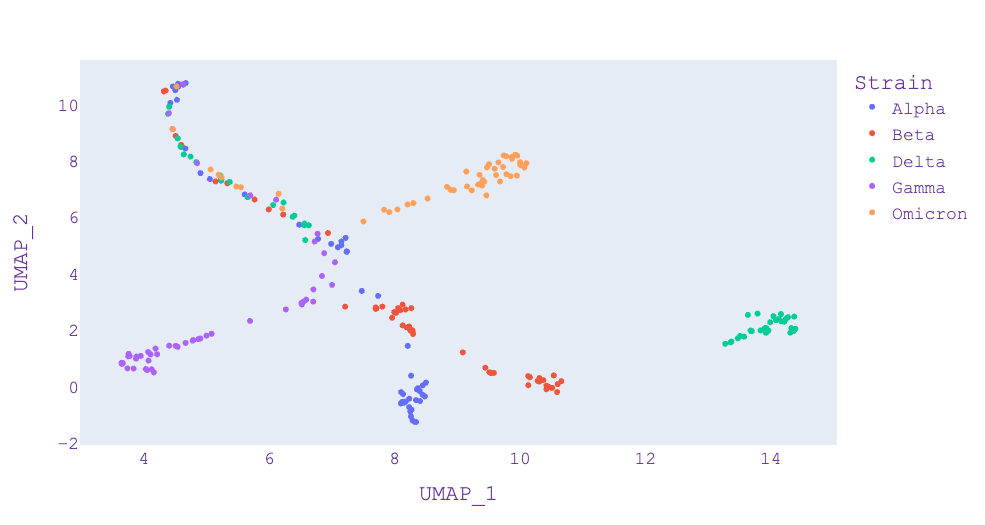}
    \vskip -0.1in
    \caption{t-SNE visualization}
    \label{fig:tsne}
\end{subfigure}

\vskip 0.1in
\caption{PCA, UMAP and t-SNE embedding visualisation from  selected SARS-CoV-2 isolates from five distinct variants using length of $k=3$.}
\label{fig:kmer3}
\end{figure}

\begin{figure}[htbp!]
\centering
\begin{subfigure}{0.75\textwidth}
    \includegraphics[width=\textwidth]{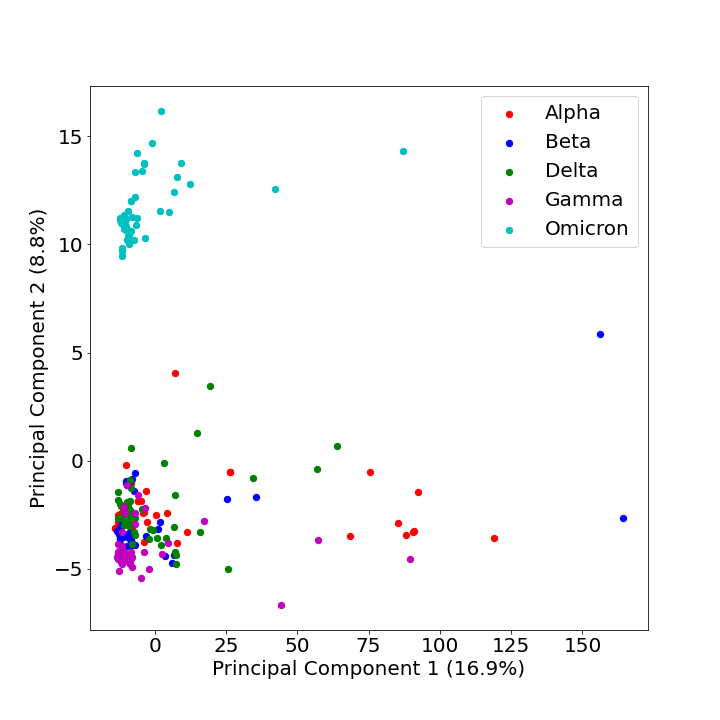}
    \caption{PCA visualization}
    \label{fig:pca_covid5}
\end{subfigure}
\hfill
\begin{subfigure}{0.75\textwidth}
    \includegraphics[width=\textwidth]{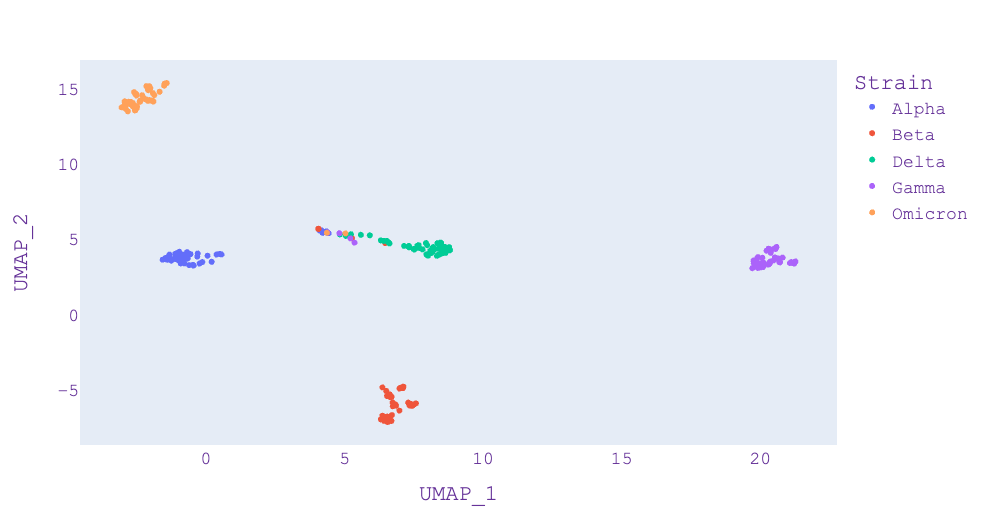}
    \caption{UMAP visualization}
    \label{fig:umap5}
\end{subfigure}
\hfill
\begin{subfigure}{0.75\textwidth}
    \includegraphics[width=\textwidth]{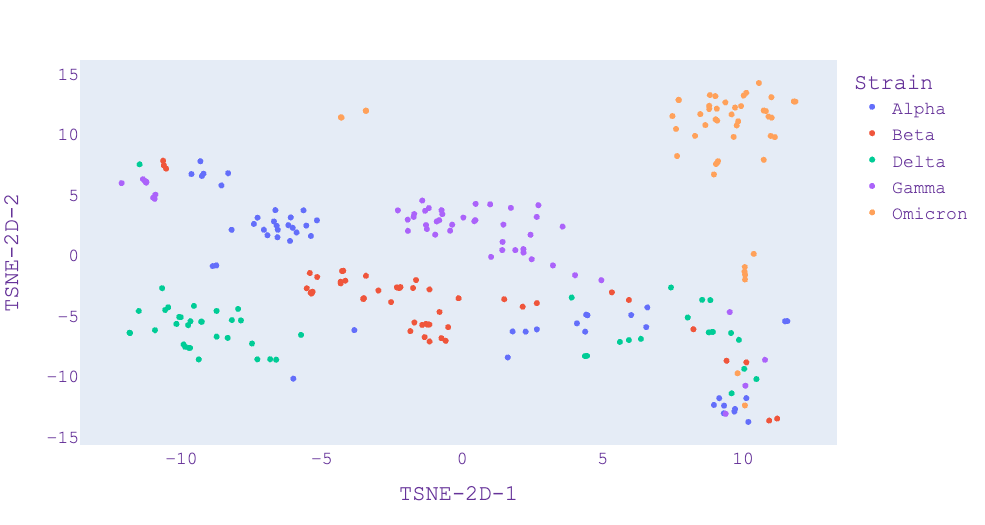}
    \caption{t-SNE visualisation}
    \label{fig:tsne5}
\end{subfigure}

\caption{ PCA, UMAP and t-SNE embedding visualisation from  selected SARS-CoV-2 isolates from five distinct variants using length of $k=5$.}
\label{fig:kmer5}
\end{figure}

\begin{figure}[htbp!]
\centering
\begin{subfigure}{0.75\textwidth}
    \includegraphics[width=\textwidth]{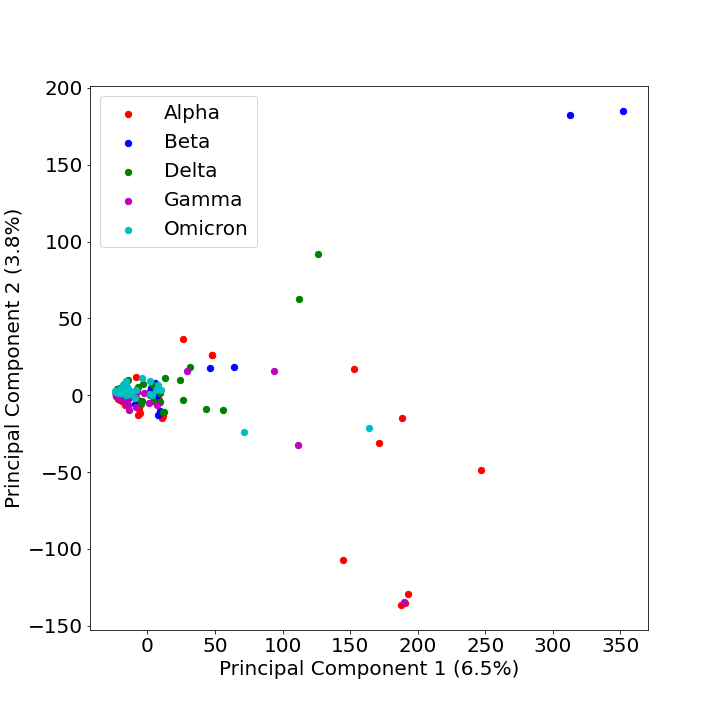}
    \caption{PCA visualization.}
    \label{fig:pca_covid7}
\end{subfigure}
\hfill
\begin{subfigure}{0.75\textwidth}
    \includegraphics[width=\textwidth]{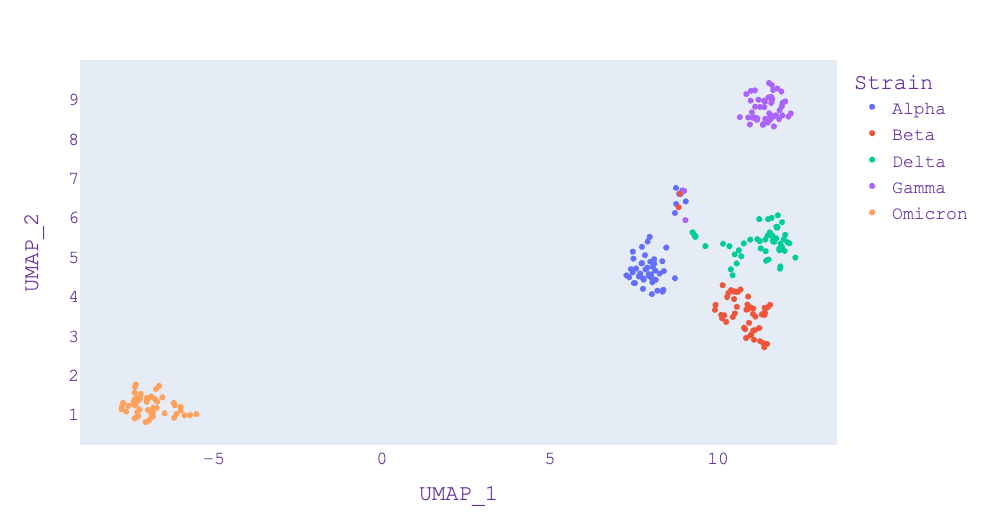}
    \caption{UMAP visualization.}
    \label{fig:umap7}
\end{subfigure}
\hfill
\begin{subfigure}{0.75\textwidth}
    \includegraphics[width=\textwidth]{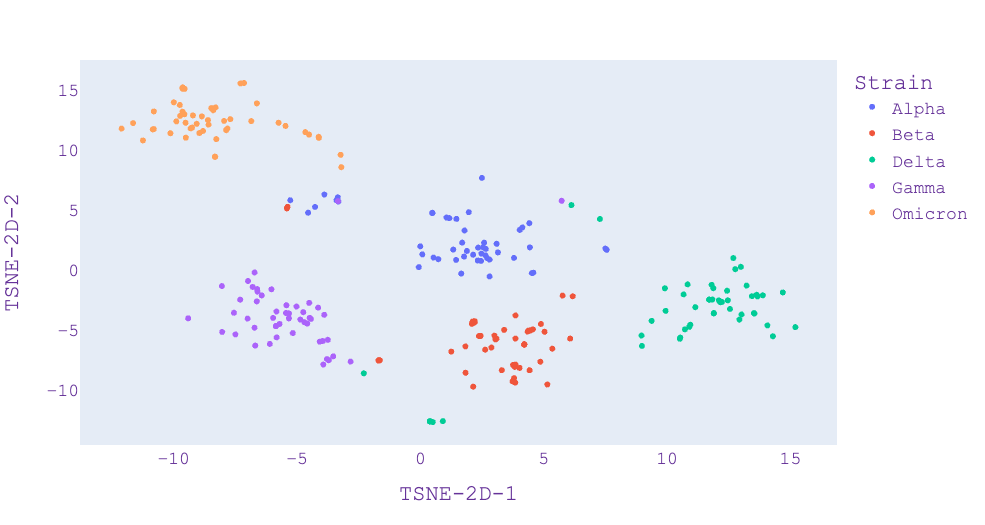}
    \caption{t-SNE visualization.}
    \label{fig:tsne7}
\end{subfigure}

\caption{ PCA, UMAP and t-SNE embedding visualisation from  selected SARS-CoV-2 isolates from five distinct variants using length of $k=7$.}
\label{fig:kmer7}
\end{figure}

\textcolor{black}{Figures \ref{fig:kmer3}, \ref{fig:kmer5} and \ref{fig:kmer7} present the visualization with PCA, UMAP, and t-SNE for selected $k$ values from $k$-mer analysis.  Figure \ref{fig:kmer3} - Panel (a) shows that   Omicron and Gamma variants are close and  overlap each other.  This is different when compared with  Figure \ref{fig:kmer5} - Panel (a) which also shows that Omicron is isolated.   Figure \ref{fig:kmer7} - Panel (a) shows that Omicron overlaps the Gamma variant. We note that with $k=5$, only 26 \% of the data is represented by the first two components (Table \ref{tab:pca}), and only 10 \%  of data is represented by the first two components of $k=7$. Hence, we can say that $k=3$ is the most reliable since it represents 61 \% of the data by the first two components. Although PCA shows a greater variance ratio for $k=3$; visually, it is poor in discriminating variants when compared to $k=5$ and $k=7$. }

In Figure \ref{fig:kmer3} - Panel (b), we find that there is a further separation of the variants using UMAP. In this case, the Alpha variant is separate while it is overlapping using PCA, as shown in Figure \ref{fig:kmer3} - Panel (a). In Figure \ref{fig:kmer3} -  Panel(c), we find that t-SNE is poor in discriminating the variants; however, t-SNE improves when $k=5$ and $k=7$ in Figure \ref{fig:kmer5} - Panel(c) and \ref{fig:kmer7} - Panel(c). In the case of UMAP, these figures show that the distance between distinct clusters becomes more apparent (increases) as the value of $k$ increases. This is also apparent in the case of t-SNE, where $k=5$ and $k=7$ provides better visualization in discriminating cluster of variants. 

Furthermore, Table \ref{tab:execution_time} presents the computational time where PCA uses the lowest computational time followed by UMAP and t-SNE. Although this is not a problem for this study since only a small dataset is utilized (250 genome sequences), the computational time would be an issue when millions of sequences need to be processed. Note that number of features obtained after the $k$-mer analysis is also shown which indicates how the dataset size changes with different values of $k$ while representing the same problem.

\begin{table}[htbp!]
    \centering
    \small
\begin{tabular}{  c c c c c  } 
%\begin{center}
 \hline
 &PCA & t-SNE & UMAP & Num. features \\
\hline
k = 3 & 0.0215  & 3.7273  & 0.2905  &  64\\ 
\hline
k = 5 &  0.0241  & 1.2987 & 0.3190 & 1024 \\ 
\hline
k = 7 & 0.2475 &  1.5757 &  0.3269 &  16384\\ 
 \hline
 %\end{center}
\end{tabular}

    \caption{Execution time (seconds) for selected values in $k$ with different number of features in data  via $k$-mer analysis.}
    \label{tab:execution_time}
\end{table}

\subsection*{Clustering}

\begin{figure*}[htbp!]
    \centering
    \includegraphics[scale = 0.3]{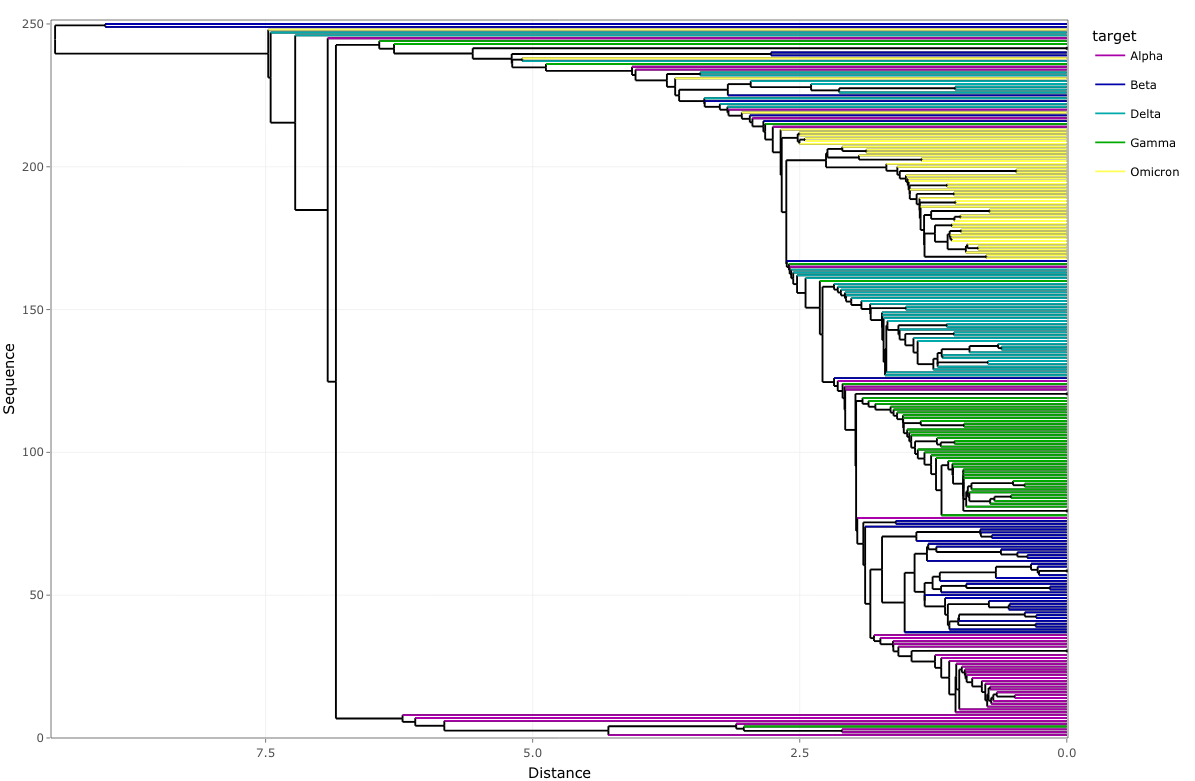}
    \caption{Dendrogram obtained from    AGNES  for  the 250 randomly extracted sequences of original data obtained after $k$-mer analysis.}
    \label{fig:dendrogram}
\end{figure*}

We apply AGNES clustering (Step 4 of the Framework in Figure \ref{fig:futurwork}) and obtain a dendrogram  using the original dataset consisting of 250 randomly selected SARS-CoV-2 genome isolates. Figure \ref{fig:dendrogram} presents the visualization obtained from the dendrogram where we can see the distinction by groups of variants.  We represent each genome isolate  by a data point using a horizontal line in the plot. The dendrogram demonstrates the relationship between   genome isolates and comprises sequences that are classified into every cluster. The value of every sequence is according to the weighted dissimilarity computation that scientists use for clustering. In Figure \ref{fig:dendrogram}, we note that  there are some mutational differences between variants of the same type owing to high mutation rates of SARS-CoV-2. These mutational differences help in identifying the variant lineage and can be used to track the route of transmission from one region to another. However, these mutations do not drastically alter variant properties. We also notice that in certain cases, certain variants such as Beta are close to Delta which is in the top cluster that falls under the distance of less than 3.0. 

\begin{figure*}[htbp!]
    \centering
    \includegraphics[scale = 0.3]{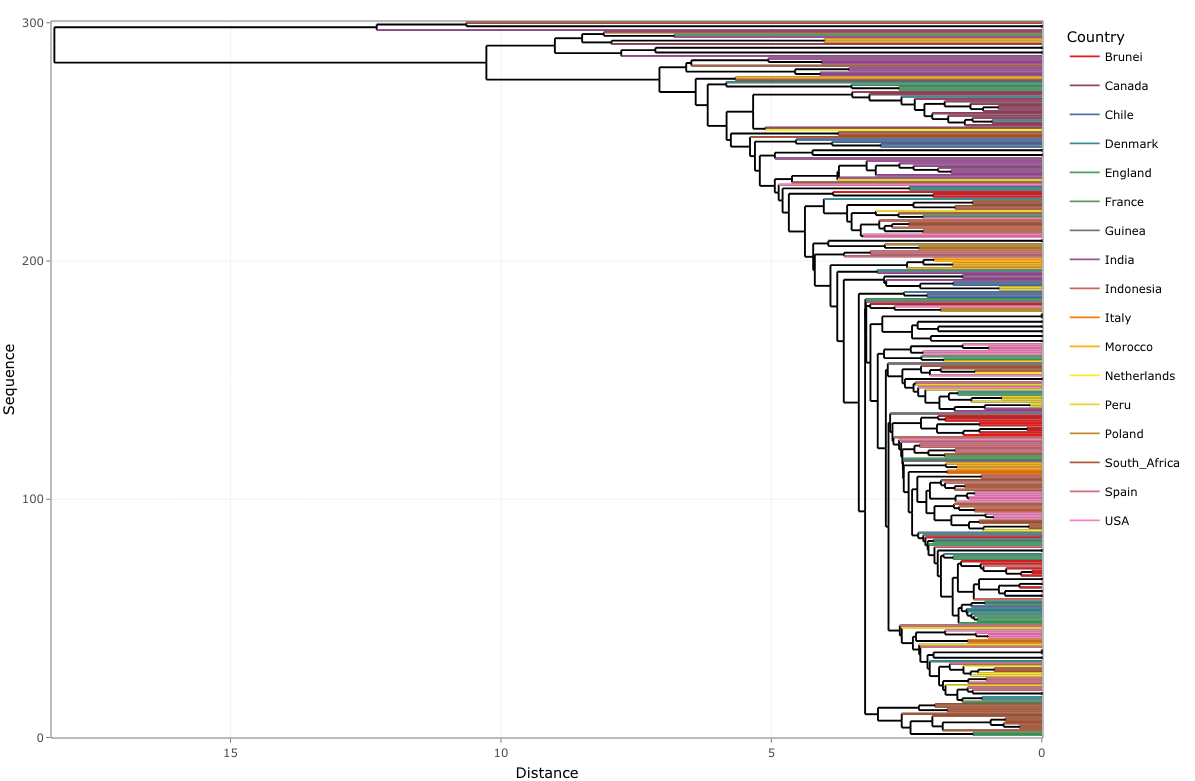}
    \caption{Dendrogram obtained from   AGNES  for  the 300 randomly extracted sequences of Omicron data obtained after $k$-mer analysis.}
    \label{fig:dendrogram_omicron}
\end{figure*}

Finally, we  apply AGNES   on a set of 300 randomly selected SARS-CoV-2 genome isolates (Table 3) of the Omicron variant  and obtain a dendrogram that shows the respective  countries and how they are related (Step 4(b)) of the Framework in Figure \ref{fig:futurwork}). \textcolor{black}{Note that we added 50 randomly selected isolates to the 250 isolates selected previously to get 300 isolates.}  Figure \ref{fig:dendrogram_omicron} presents the dendrogram obtained from the visualization for Omicron where we can see that there are some mutational differences between Omicron genome sequences from different regions. As mentioned above, these give rise to different lineages but do not alter the viral phenotype. We also observe that there is a certain level of similarity between variant sequences from different regions, for example, (USA, Spain, Brunei) and (India, Morocco, South Africa). In future work, the point of origin of these similarities can be traced by doing a spatiotemporal analysis of the data. Hence, once extended, this framework has the potential to  track the trend of viral spread from one region to another.

\begin{figure*}[htbp!]
    \centering
    \includegraphics[scale = 0.3]{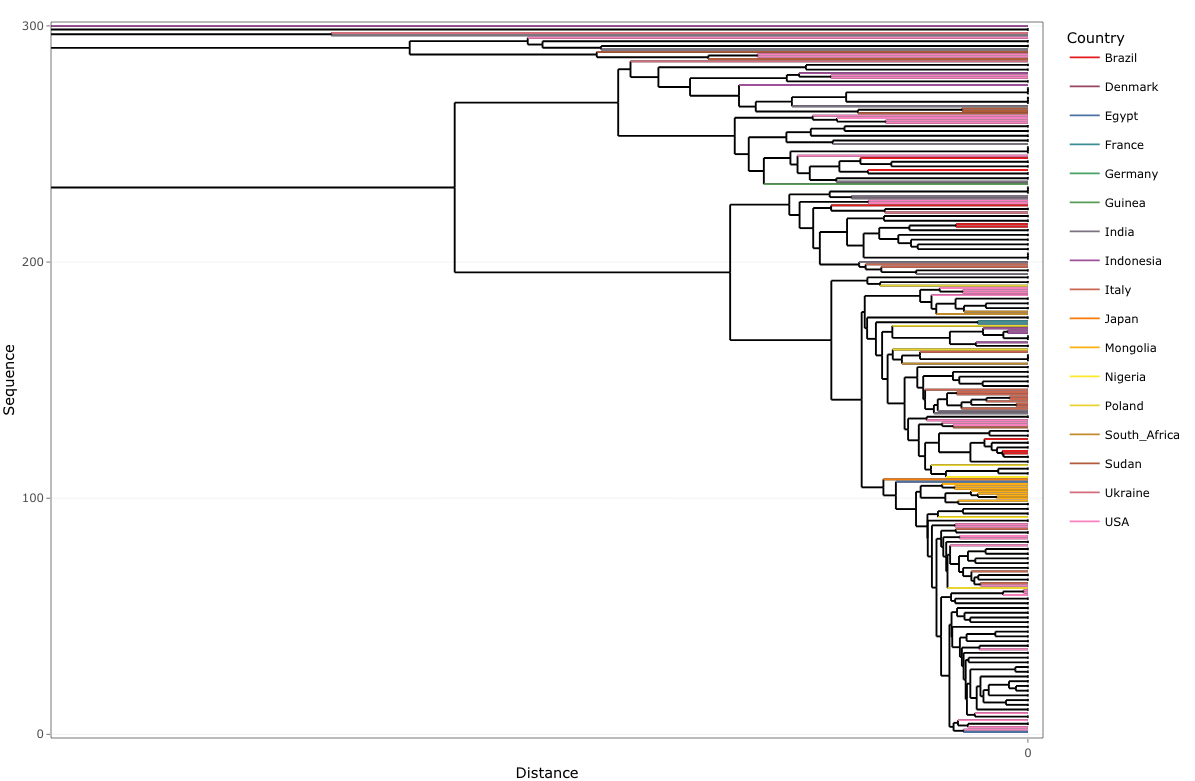}
    \caption{Dendrogram obtained from   AGNES  for  the 300 randomly extracted sequences of Delta data obtained after $k$-mer analysis.}
    \label{fig:dendrogram_delta}
\end{figure*}

A similar country-wise analysis was also performed on 300 randomly selected SARS-CoV-2 genome isolates of the Delta variant. Figure  \ref{fig:dendrogram_delta} presents the dendrogram obtained from the visualization for Delta, where we observe that the level of similarity for the genome sequence among different regions was much higher than that observed for Omicron. This supports the fact that the observed number of lineages for Omicron(7) is more than that for Delta(2) \cite{varInfo}, which further hints at a higher mutation rate for the Omicron variant. 

\section*{Discussion}

The main contribution of this study  lies in examining the COVID-19 isolates using classical and  novel dimensionality reduction and clustering methods. 
In general, we found  that UMAP performs better  than PCA and t-SNE for the given COVID-19 genome isolates. It is effective in visualizing clusters as it takes the nonlinearity of the data into account, unlike PCA, and can capture the global structure of the data better than t-SNE. We also note that $k$-mer analysis is an important data pre-processing step when dealing with genomics data. Our results show that the value of $k$ plays a crucial role in capturing the features.  Depending on the method (PCA, UMAP, and t-SNE), it is critical to choose the right value of $k$ for $k$-mer analysis. Even though small values of $k$ often lead to information loss, a larger value of $k$, while preserving important information, demands more computational resources. Thus, we conclude that it is reasonable not to go further than $k=7$, as it can take further computational time  (Table 5) and storage during genome sequence pre-processing. \textcolor{black}{We note that $k=3$ shows the highest explained variance ratio of PCA in Table 4. Figure 3 shows that $k=3$ from PCA  is not as good as  UMAP which has been better in discriminating the variants. Hence, although $k=3$ has a high explained variance ratio, it is not  appropriate when we check the visualisation and  compare it with $k=7$ in Figure 5 where  UMAP shows the best results. Hence, we can note that PCA is not an appropriate method for visualisation in this case and $k=7$ is the best $k$-mer count value which is supported by UMAP results.}

UMAP is a nonlinear dimensionality reduction method that creates \textit{simplicial complexes} by connecting points if the distance between them is below a threshold. UMAP uses these complexes to calculate the relative distance in the lower dimensions, unlike t-SNE, which does it randomly \cite{mcinnes2018umap}. PCA on the other hand, cannot capture non-linear dependencies as it is a linear projection and its primary goal is to find directions that maximize the variance in the dataset. Due to these reasons, UMAP scales well given different variations in $k$-mer analysis and also provides a better visual representation with less computational time when compared with PCA and t-SNE (Figures 3, 4 and 5). On the other hand, PCA provides further insights using explained variance ratio which in addition while  UMAP gives a good overview of the data. It will be useful to have the feature of explained variance ratio in UMAP and t-SNE, this can shed more light on $k$-mer analysis   as done by PCA. 

In our framework,  we utilise dendrograms via AGNES  for visualizing the mutational differences and similarities among various groups. AGNES is easy to implement as it does not require prior information about the number of clusters but the time complexity is high and thus it is computationally expensive for larger and more complex datasets. Similarly, dendrograms well interpret but become less resourceful as the complexity of data increases. 

 Phylogeny reconstruction   describes evolutionary relationships in
terms of the \textit{relative recency} of common ancestry which are typically represented as a branching diagram, or
tree, with branches joined by nodes \cite{jill2006step}. Phylogeny reconstruction has been prominent in studying evolutionary biology, particularly in visualising the relationship amongst genome sequences \cite{boore2006use}; and hence, it has been used for COVID-19 isolates \cite{taiwo2022sequence}. Our framework produces dendograms that are similar to phylogeny reconstruction, with additional visualisation using dimensional reduction methods.  Phylogeny reconstruction has certain limitations which have been demonstrated for amido acid sequence data \cite{bremer1988limits}.  The phylogeny trees may not  necessarily accurately represent the evolutionary history of the data \cite{hoelzer1994patterns}. A dendrogram is a general term for    any type of phylogeny tree (scaled or unscaled).  Hence, our framework is not replacing phylogeny reconstruction, but providing a means to provide additional information. We not only provide the dendrograms but also visualization by dimensionality reduction methods to distinguish variants of concern. In future work, a comparison between dendrogram and phylogeny reconstruction  would highlight the strengths and limitations of the different approaches.

In future work, the proposed framework can be extended further with novel  dimensionality reduction and clustering methods.  Therefore, the   other novel dimensionality reduction approaches such as  \textit{Ivis}  \cite{ivis}  could be considered which is good in  extremely large datasets. The genome data extraction using $k$-mer analysis can be compared with alternatives, such as \textit{strobemers} \cite{sahlin2021effective,sahlin2021strobemers} which is gaining attention in the area of genome sequence analysis.

  We note that given the efficacy of the framework in distinguishing different variants of concern, the framework can be used for  assisting scientists and policymakers in pandemic management. The  framework can be used for large-scale temporal and spatial study of the emergence of major variants of COVID-19 in selected countries, and also globally which can help in better understanding the infection and death rate trend. This can also give an insight into the effectiveness of vaccination programs and boosters \cite{vaccine_variant} for different variants. Furthermore, the framework can be used to perform a spatiotemporal analysis to study the pattern of the spread of infection from one region to another. It can also be extended to perform a similar analysis on future outbreaks (pandemics)  to understand  the nature of  emerging variants. Finally,  the web-based application can be developed using our framework that features geo-location and interactive maps (country-wise and worldwide) displaying different variants and their evolution over time.

The limitations of the study include the meta-information provided in the COVID-19 genome isolates since a large number of samples only have dates associated with the data uploaded rather than when they were taken. It is important to know meta-information such as the date and time of the samples collected in order to have further insights into the changing nature of the variants. We also need to note that the number of variants and the number of samples for each variant are magnitudes lower than the number registered in the dataset. Our framework handled a few hundred samples and could be extended to thousands of samples. However, catering millions of samples across space (geo-location) and time would be computationally intensive and parallel computing facilities would be needed. 

\section*{Conclusion}

We presented a framework that provides  insights that can further help scientists in effectively discriminating the COVID-19 variants that rapidly change due to mutations. In our framework, we evaluated the dimensional reduction components of the framework with different methods and found that UMAP provides the best dimensionality  reduction and visualization tool for the genome sequences. We showed that PCA used in conjunction with t-SNE and UMAP addresses the limitations of the latter methods since they do not provide explained variance ratio. In many applications, visualization of the data alone is not sufficient to address the problem; it is critical to know the contribution of the different features which can only be known through PCA. Furthermore, the visualization of the emerging COVID-19 variants using dendrograms via clustering can provide detailed insights about their evolution which can be extended to larger datasets.

%\section*{Supporting information}
%\subsection*{Code and Data }
%Open-source Python and R code for the framework is available on GitHub:
%\url{https://github.com/ai-covariants/analysis-mutations}.

%  
%\bibliographystyle{plos2015} 
% \bibliography{cas-refs,sample}

\end{document}